\documentclass[aps,pre,twocolumn,showpacs,superscriptaddress,
amsmath,amssymb]{revtex4}
\usepackage{graphicx}
\usepackage{dcolumn}
\usepackage{bm}
\begin{document}
\title{Dipolar origin of the gas-liquid coexistence of the hard-core 1:1
electrolyte model}
\author{J. M. Romero-Enrique}\email[Corresponding author. 
Electronic address: ]{enrome@us.es}
\affiliation{Departamento de F\'{\i}sica At\'omica, Molecular
y Nuclear, \'Area de F\'{\i}sica Te\'orica, Universidad de Sevilla, Aptdo.
Correos 1065, 41080, Spain.}
\author{L. F. Rull}
\affiliation{Departamento de F\'{\i}sica At\'omica, Molecular
y Nuclear, \'Area de F\'{\i}sica Te\'orica, Universidad de Sevilla, Aptdo.
Correos 1065, 41080, Spain.}
\author{A. Z. Panagiotopoulos}
\affiliation{Department of Chemical Engineering, Princeton University,
Princeton NJ 08544, USA}
\date{\today}
\begin{abstract}
We present a systematic study of the effect of the ion pairing on
the gas-liquid phase transition of hard-core 1:1 electrolyte
models. We study a class of dipolar dimer models that depend on a
parameter $R_c$, the maximum separation between the ions that
compose the dimer. This parameter can vary from $\sigma_\pm$ that
corresponds to the tightly tethered dipolar dimer model, to
$R_c\to \infty$, that corresponds to the Stillinger-Lovett
description of the free ion system. The coexistence curve and
critical point parameters are obtained as a function of $R_c$ by
grand canonical Monte Carlo techniques. Our results show that this
dependence is smooth but non-monotonic and converges
asymptotically towards the free ion case for relatively small
values of $R_c$. This fact allows us to describe the gas-liquid
transition in the free ion model as a transition between two
dimerized fluid phases.  The role of the unpaired ions can be
considered as a perturbation of this picture.
\end{abstract}
\pacs{64.70.Fx, 05.10.Ln, 05.70.Jk}
\maketitle

\section{Introduction}

In recent years there has been an increasing interest in phase
transitions between fluid phases of different electrolyte
concentrations in ionic solutions. Two different regimes have been
identified experimentally \cite{Fisher-review}. The
``solvophobic'' regime occurs for large solvent dielectric
constants that effectively turn off Coulombic interactions.
Consequently, solvophobic phase transitions are primarily driven
by unfavourable interactions between solute and solvent. This
behavior is well described by the usual theory of nonelectrolyte
solutions with short range interactions, and clearly leads to
Ising-like critical behavior. By contrast, in the ``Coulombic''
regime the solvent has a low dielectric constant and electrostatic
interactions between the solute ions drive the phase separation.
In this case, the phase diagrams are quite asymmetric, and
apparently mean-field critical behavior has been claimed
\cite{Singh,Weingatner,Zhang}, although Ising-mean field crossover
has also been seen within a narrower range of temperatures around
the critical than in non-ionic fluids
\cite{Narayanan,Jacob,Anisimov,Gutkowski}.  Some of these
experimental studies suggest that there exists a new
characteristic length in these systems that competes with the
correlation length for density fluctuations \cite{Anisimov,
Gutkowski}.

Electrolyte systems in the Coulomb regime are often modelled as
charged hard spheres embedded in a uniform dielectric continuum
(primitive models). Most studied is the ``restricted primitive
model'' (RPM), in which the ions are of equal size. A vapor-liquid
phase transition at very low temperatures and densities was
predicted theoretically 25 years ago \cite{Stell1} and by early
computer simulation studies \cite{Vorontsov}. Improvement of
computer simulation techniques has allowed an increasingly precise
determination of the coexistence parameters
\cite{Orkoulas,Caillol,DePablo1,Thanos,Thanos4,Luijten}.  There
have also been recent studies of primitive models with asymmetry
in size \cite{E2PM,DePablo2,Thanos2} and charge
\cite{Thanos2,DePablo3}.  Very recent results suggest that the
critical behavior of the RPM belongs to the Ising universality
class \cite{Luijten}.

From a theoretical point of view, different approaches have been
used in order to explain the vapor-liquid transition of the
primitive models. Integral equations such as the mean spherical
approximation (MSA) \cite{Levin,Tovar}, as well as the
Debye-H\"uckel theory \cite{Levin} and Poisson-Boltzmann
approaches \cite{Outhwaite} have succesfully been applied to it.
For the RPM, the most succesful theories are the pairing theories,
that consider the ionic fluid as a mixture of bound pairs and free
ions in chemical equilibrium \cite{Levin,Raineri}, in the spirit
of the Bjerrum's ideas \cite{Bjerrum}.  However, in all these
theories the transition is driven by the free ions, even when in
same cases, as in the Debye-H\"uckel-Bjerrum approach, the
associated pairs are the dominant species.  Analytical
\cite{Gillan} and computer simulation \cite{Bresme} results, on
the other hand, show that the structure of the vapor phase is
dominated by neutral clusters, mostly dimers and tetramers.
Computer simulations have also demonstrated that the phase
envelope of the RPM resembles that of the charged hard dumbbell
model \cite{Shelley,E2PM}. This result has been confirmed by
theoretical studies \cite{Kalyuzhnyi,Jiang}.  The question that
the present work examines in detail is the influence of ionic
association on the vapor-liquid transition of primitive model
electrolytes.  In contrast to an earlier study \cite{E2PM}, which
only considered tightly bound dimers, here we examine a range of
models with varying values of $R_c$, the maximum separation
between ions in a dimer.  $R_c\to \infty$ corresponds to the
Stillinger-Lovett description \cite{Stillinger} of the free ion
system.

The structure of this paper is as follows.  In Section \ref{sec2},
we examine the microscopic structure of the coexisting phases of
the ionic fluid and compare the correlation functions to those of
a tightly tethered dimer model. In Section \ref{sec3}, an exact
chemical representation of the ionic fluid as a mixture of
associated pairs and free ions is introduced.  The role of the
associated pairs on the phase coexistence is studied in detail in
Section \ref{sec4} and the paper closes with discussion and
conclusions.

\section{Microscopic structure of the ionic and tethered dimer systems at coexistence}
\label{sec2}
 In this Section we analyse the role of pairing on
gas-liquid coexistence of primitive model ionic fluids.  We have
studied by computer simulation an 1:1 size-asymmetric primitive
model, in which the ions are modelled as hard spheres of diameters
$\sigma_+$ and $\sigma_-$, and carrying charges $+q$ and $-q$,
respectively, embedded on a dielectric continuum of dielectric
constant $D$ ($D=1$ for the vacuum). The interaction potential
between two ions separated by a distance $r_{ij}$ is given by:
\begin{equation}
u_{ij}(r_{ij})=u_{hs}(r_{ij})+\frac{q_i q_j}{D r_{ij}}
\end{equation}
where $u_{hs}(r_{ij})$ is the hard-core potential that takes the value
$+\infty$ if $r_{ij}<\frac{1}{2} \left(\sigma_i + \sigma_j\right)$ and 0
otherwise.

The size asymmetry is characterized by the parameter $\lambda$, defined as:
\begin{equation}
\lambda=\frac{\sigma_-}{\sigma_+}
\end{equation}

Monte Carlo simulations in the neutral grand-canonical ensemble were
performed, characterized by a temperature $T$ and the
\emph{configurational} chemical potential for a pair of unlike ions
$\mu=2\mu_{real}-3k_B T \ln \left(\Lambda_+ \Lambda_-\right)$,
with $\mu_{real}\equiv(\mu_+ + \mu_-)/2$ and $\Lambda_\pm$ are the
thermal de Broglie wavelengths of the ionic species:
\begin{equation}
\Lambda_{\pm}=\sqrt\frac{h^2}{2\pi m_{\pm} k_B T}
\label{broglie}
\end{equation}
As usual, cubic boxes of length $L$ under periodic boundary
conditions are used. The long-ranged character of the Coulombic
interactions is handled by the use the Ewald summation tecnique
with conducting boundary conditions, with 518 Fourier-space
wavevectors and real-space damping parameter $\kappa=5$. The
relative error due to the infinite sums truncation in the
electrostatic energy is less than $10^{-5}$ for random
configurations in small systems \cite{Hummer}, and this choice has
been also validated by direct simulations of the RPM
\cite{Thanos4}. In order to speed up the simulations, the basic
steps of the simulations (insertion and deletions of pairs of
unlike ions) are biased following \cite{Orkoulas}. Moreover, a
fine-discretization approximation is used: the positions available
to each ion are the sites of a simple cubic lattice of spacing
$a$. This methodology has succesfully been applied to the RPM
\cite{Thanos}, to 1:1 size-asymetric primitive models \cite{E2PM}
and to $z$:1 size-asymetric primitive models \cite{Thanos}, and
allows a speedup relative to the continuum calculations of a
factor of 100 for small systems. The results are almost
indistinguishable from the continuum ones for a discretization
parameter $\zeta\equiv \sigma_\pm/a=10$ \cite{Thanos,E2PM,
Thanos2}, where $\sigma_\pm=\frac{1}{2}\left(\sigma_+ +
\sigma_-\right)$ is the unlike-ion collision diameter. This value
of the discretization parameter ($\zeta=10$) was used in the
present study.  Histogram reweighting techniques \cite{Ferrenberg}
and mixed-field finite size scaling methods \cite{Bruce} were used
to obtain the vapor-liquid envelopes and the effective critical
points, respectively. For the sake of comparison, we have also
studied tightly tethered dipolar dimer systems
\cite{Shelley,E2PM}, consisting of $N$ pairs of a positive and
negative ion restricted to remain at separations $\sigma_d$
satisfying $\sigma_\pm \le \sigma_d \le 1.02\sigma_\pm$.
Simulation details and some preliminary results for a range of
values of $\lambda$ are presented in Ref. \cite{E2PM}.

The unlike-ion collision diameter $\sigma_\pm$ provides the basic length scale
appropriate for defining both the reduced temperature and reduced density via
\begin{equation}
T^*=k_B T D \sigma_\pm/q^2\quad \mathrm{and}\quad \rho^*=2 N \sigma_\pm^3
/L^3
\end{equation}
where $N_+=N_-\equiv N$ is the particle number of each ionic species
\cite{Tovar,Orkoulas,E2PM}. The reduced simulation box length is defined
similarly via $L^*=L/\sigma_\pm$, and the reduced energies and chemical
potential as $U^*=UD\sigma_\pm/q^2$ and $\mu^*=\mu D\sigma_\pm/q^2$.

\begin{figure}
\includegraphics[width=8.5cm]{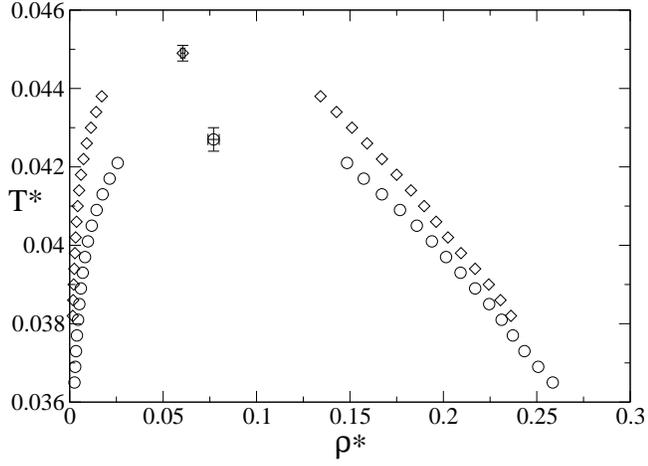}
\caption{Gas-liquid phase diagram for the ionic system (diamonds)
and the tethered dimer fluid (circles) with a tether length equal to
$1.02\sigma_\pm$. The size asymmetry parameter is set $\lambda=3$. The critical
parameters have been obtained from the extrapolation to $L^*\to \infty$
of the results for $L^*=12,15$ and $18$ while the subcritical coexistence
curves were obtained using $L^*=15$.\label{fig1}}
\end{figure}

The value of the asymmetry parameter considered in this paper
is $\lambda=3$. For that case, the gas-liquid coexistence of the ionic
fluid shows a shift in both temperature and density respect to the
tethered dimer fluid (see Fig. \ref{fig1}), which is a general feature when
comparing ionic and tethered dimer systems \cite{E2PM}. On the other hand
the asymmetry is not high so as to favor large chain-like neutral clusters,
as occurs for bigger values of $\lambda$ \cite{E2PM,DePablo2}.
These features qualify this case to be a typical example for moderately
asymmetric 1:1 electrolytes, including the RPM.

\begin{figure}
\includegraphics[width=8.5cm]{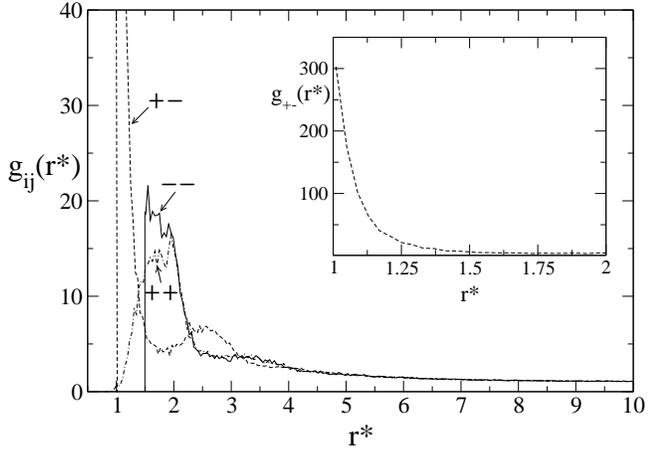}
\caption{Ion-ion radial distribution functions corresponding to the ionic
system at $T^*=0.0425$ and $\rho^*=0.0076(15)$ (the gas phase).
In the inset, the behavior of $g_{+-}(r^*)$ at distances close to
the contact value.\label{fig2}}
\end{figure}

Fig. \ref{fig1} makes the similarity between the phase diagram
of the ionic and the tethered dimer fluids clear. It also suggests
in an indirect way that the pairing plays a decisive role on the
gas-liquid coexistence in the
ionic fluid. In order to clarify such a role, we have studied the ion-ion
radial distribution functions $g_{ij}(r)$ corresponding to the ionic
systems, and the corresponding ones for the tethered dimer systems.
For this purpose, we have considered the gas and liquid states at coexistence
for a temperature $T\approx 0.95 T_c$, with $T_c$ the corresponding critical
temperature. The ion-ion radial distribution functions of the ionic systems,
for $T^*=0.0425$ and the coexisting densities $\rho^*=0.0076(15)$ and
$\rho^*=0.162(3)$ are plotted in Figs. \ref{fig2} and \ref{fig3},
respectively. In both cases, the unlike-ion radial distribution function
becomes very large close to contact, indicating the association in bound pairs
of unlike ions. Moreover, the like-ion radial distribution functions show
maxima around $r^*=1.5$ (in the case of $g_{--}$, this peak coincides
with the contact value), and there is a secondary maximum in $g_{+-}$
for $r^*\approx 2.5$. These observations allow us to conclude that there
are high correlation between pairs of associated unlike ions.
We recall that the range of densities in which the
gas-liquid coexistence occurs prevents packing effects, so the structure
is completely given by the Coulombic interactions.

The ion-ion radial distribution functions for the tethered dimer
fluid are plotted on the Fig. \ref{fig4} for the gas branch
($T^*=0.0405$, $\rho^*=0.0119(15)$) and Fig. \ref{fig5} for the
liquid branch ($T^*=0.0405$, $\rho^*=0.206(2)$). The comparison
between the ionic fluid and tethered dimer fluid microscopic
structures confirms the qualitative similarity between both
systems. A further test of this similarity is found in the
comparison of the neutral cluster populations in the gas branch.
We use Gillan's definition of a cluster \cite{Gillan}. Two ions
$i$ and $j$ are directly bound when the distance between them is
less than $R_{ij}^c$. This condition defines mathematically an
equivalence relationship, and the equivalences classes in which
the ions group are the clusters. As the interaction between like
ions is repulsive, the dependence of the cluster definition on
$R_{++}^c$ or $R_{--}^c$ should be very weak (if they are taken
smaller than the mean distance between two like ions). On the
other hand, the cluster definition is not very sensitive to the
value of $R_{+-}^c$ if that value lies between the first minimum
of $g_{+-}$ and the mean distance between aggregates. In this work
we have used $R_{++}^c=R_{--}^c=R_{+-}^c=1.5\sigma_\pm$. As
expected, the microscopic structure of the gas phase is dominated
by the $N$-ion neutral clusters \cite{Bresme,E2PM}. Their
fractions $f_N$ for the ionic and tethered ion systems at the gas
phase in coexistence for $T=0.95 T_c$ are quite similar (see Fig.
\ref{fig6}), although the tethered dimer systems have slightly
higher fractions of neutral clusters than do ionic systems,
specially for large $N$, in agreement with the results previously
reported \cite{E2PM}.

Despite the similarities found between the ionic fluid and the
tethered dimer fluid, there are some differences that can rationalize
the quantitative differences between their phase diagrams. The unlike-ion
radial distribution functions of the tethered dimer fluid differ
qualitatively from the ionic counterparts close to the contact value, since
in the former there is a jump at the maximum allowed separation of two ions
of a dimer, while the ionic $g_{+-}$ takes smoothly higher values
on that range of values of $r^*$. Consequently, the condition on the
confinement of the ions that compose a tethered dimer must be relaxed
in order to refine the pairing concept in the ionic fluids. In the next
Section we will introduce an appropriate framework for such a goal.

\section{Chemical picture of ionic fluids}
\label{sec3} We consider an ionic fluid ($q_+=-q_-\equiv q$) be
contained in a volume $V$ and in equilibrium with a reservoir at a
temperature $T$ and chemical potentials $\mu_+$ and $\mu_-$. For
simplicity, the neutral grand-canonical ensemble, in which only
neutral configurations are allowed, will be considered. This
ensemble has been shown to be equivalent to the usual
grand-canonical ensemble in the thermodynamic limit \cite{Lieb}.
The neutral grand-canonical partition function can be written as:
\begin{equation}
\Xi=\sum_{N_i=0}^\infty \left(\frac{\textrm{e}^{\beta \mu}}{\Lambda_+^3}
\right)^{N_i}\left(\frac{\textrm{e}^{\beta \mu}}{\Lambda_-^3}\right)^{N_i}
Z_U(N_i,N_i)
\label{partition}
\end{equation}
where $N_i$ is the number of ions of each species, $\beta=1/(k_B T)$,
$\mu\equiv(\mu_+ + \mu_-)/2$, $\Lambda_+$ and $\Lambda_-$
are the thermal de Broglie wavelengths corresponding to each species,
and $Z_U(N_i,N_i)$ is the canonical configurational partition function
corresponding to a fixed number of ions at the temperature $T$ and enclosed
in the volumen $V$:
\begin{equation}
Z_U(N_i,N_i)=\frac{1}{(N_i !)^2}
\int_{V^{2 N_i}}d\mathbf{r}^+_1 d\mathbf{r}^-_1 \ldots
d\mathbf{r}^+_{N_i}d\mathbf{r}^-_{N_i} \textrm{e}^{-\beta U}
\label{partition2}
\end{equation}
with $U\equiv U^{phys}= U_{hc}+\sum_{i<j} q_i q_j/(D r_{ij})$ the
total potential energy and $U_{hc}$ the hard-core contribution,
equal to $+\infty$ if two particles overlap, and $0$ otherwise
(other tempered potentials can be used, but the main results of
this Section will remain unchanged). If the fluid particles are
strongly associated into bound $(+,-)$ pairs, as occurs in the
low-temperature and low-density region in which the vapor-liquid
transition occurs for the ionic fluids, the ``physical''
representation described above will be inadequate, and it can
replaced by a ``chemical'' picture, in which the fluid is composed
by associated pairs and free ions. First, a rule that
unequivocally identifies bound pairs for each configuration of the
ionic fluid (up to a set of null measure in the configurational
space) is needed.  Once such a rule is defined, we can write the
canonical configurational partition function as:
\begin{equation}
Z_U(N_i,N_i)=\sum_{N_p=0}^{N_i} Z_U^*(N_f,N_f,N_p)
\label{decompose}
\end{equation}
where $N_p$ is the number of bound pairs, $N_f=N_i-N_p$ is the
number of free ions of each species, and $Z_U^*(N_f,N_f,N_p)$ is
the canonical configurational partition function corresponding to
a system of $N_p$ associated pairs and $N_f$ free ions of each
species:
\begin{eqnarray}
Z_U^*(N_f,N_f,N_p)&=&\frac{1}{N_p!} \frac{1}{(N_f !)^2}
\int_{V^{2(N_p+N_f)}}d(1)\ldots d(N_p)\nonumber\\
&\times &d\mathbf{r}^+_1 d\mathbf{r}^-_1 \ldots
d\mathbf{r}^+_{N_f}d\mathbf{r}^-_{N_f} \textrm{e}^{-\beta U^*}
\label{effectiveham}
\end{eqnarray}
where $(i)\equiv\{\mathbf{r}^+,\mathbf{r}^-\}_i$ are the positions of the ions
that compose the associated pair $i$, and ${\bf r}^{+(-)}_i$ corresponds to
the coordinates of a $+(-)$ free ion $i$, respectively. The potential energy
$U^*\equiv U^{chem}$ \emph{does not} coincide, in general, with the physical
potential energy $U^{phys}$, since different configurations of the ``chemical''
mixture can be compatible with a given ionic configuration.

Substituting equation (\ref{decompose}) into the equation
(\ref{partition}) and rearranging the resulting expression, the
neutral grand-canonical partition function can be written as:
\begin{eqnarray}
\Xi&=&\sum_{N_f=0}^\infty \sum_{N_p=0}^\infty
\left(\frac{\textrm{e}^{\beta \mu}}{\Lambda_+^3}
\right)^{N_f}\left(\frac{\textrm{e}^{\beta \mu}}{\Lambda_-^3}\right)^{N_f}
\left(\frac{\textrm{e}^{\beta \mu_p}}{\Lambda_p^6}\right)^{N_p}\nonumber\\
&\times& Z_U^*(N_f,N_f,N_p)
\label{partition3}
\end{eqnarray}
with $\mu_p\equiv 2\mu$ and $\Lambda_p=\sqrt{\Lambda_+
\Lambda_-}$. These last definitions correspond to the chemical
equilibrium conditions between the free ions and the associated
pairs \cite{Chandler,Levin}. We remark that this derivation is
\emph{independent} of the criterion chosen to define the pairs.
However, the choice must be such that matches the microscopic
structure of the fluid. As we have seen in the previous Section,
the vapor structure is dominated by neutral aggregates, mostly
dimers and tetramers. This fact is consistent with the Bjerrum
ideas of pairing \cite{Bjerrum}. Consequently, a distance-based
criterion seems to be the most appropriate: two unlike ions that
are closer than $R_c$ ($R_c$ being a suitable cutoff distance) are
considered as an associated pair.   It is easy to see that such a
criterion \emph{does not} define uniquely the associated pairs
when the population of tetramers and higher order neutral clusters
is not negligible, as it can be seen from Fig. \ref{fig7}, since
there are different arrangements of associated pairs that are
compatible with the same ionic configuration.  Hence, a more
systematic criterion is needed, in order to be able to define
associated pairs for \emph{almost every} ionic configuration,
while keeping the intuitive definition of an associated pair as
encompassing unlike ions that are at the closest distances. We use
a suitable modification of the Stillinger-Lovett pair definition
on a given ionic configuration \cite{Stillinger}. In this
prescription, all the distances between two unlike ions are
computed, and then the first pair is \emph{defined} as the two
unlike ions at closest distance. This step is repeated, taking
into account only ions that remain unpaired from previous steps
for the evaluation of $(+,-)$ distances. We stop when the minimum
distance between two unlike ions is greater than $R_c$,
considering the remaining ions as free ions (no free ions were
considered in \cite{Stillinger}, and consequently $R_c \equiv
\infty$ in that case). This protocol provides an unique
configuration of associated pairs and free ions for almost each
ionic configuration, since the method is ambiguous only in a
subset of the ionic configurations in which at least two $(+,-)$
distances are equal, and the measure of such a set is null in the
configurational space.

Obviously, different pairing prescriptions can be used. However, this protocol
provides an explicit expression for the ``chemical'' potential energy
$U^{chem}$:
\begin{widetext}
\begin{eqnarray}
U^{chem} &=& U^{phys}+
\sum_{i=1}^{N_p} v^p(\{{\bf r}^+,{\bf r}^- \}_i)
+\sum_{1\le i < j \le N_p} v^{pp}(\{{\bf r}^+,
{\bf r}^- \}_i,\{{\bf r}^+,{\bf r}^- \}_j)
\nonumber\\
&+&\sum_{i=1}^{N_p}\sum_{j=1}^{N_f} v^{p+}(\{{\bf r}^+,{\bf r}^- \}_i,
{\bf r}^+_j)
+ \sum_{i=1}^{N_p}\sum_{j=1}^{N_f} v^{p-}(\{{\bf r}^+,{\bf r}^- \}_i,
{\bf r}^-_j)
+\sum_{i=1}^{N_f} \sum_{j=1}^{N_f} v^{+-}({\bf r}^+_i,{\bf r}^-_j)
\label{hameffec}
\end{eqnarray}
\end{widetext}
where $U^{phys}$ is the physical potential energy due to the
hard-core and eletrostatic interactions, and the other terms correspond to
effective interactions of entropic origin needed to reduce the
configurational space of the ``chemical'' system.
The term $v^p(\{{\bf r}^+,{\bf r}^- \}_i)$ depends
only on the positions of the ions that compose an associated pair and confines
them to be at a distance shorter than $R_c$:
\begin{equation}
v^p(\{{\bf r}^+,{\bf r}^- \}_i) = \begin{cases}
0 & |{\bf r}^+_i-{\bf r}^-_i|\le R_c \\ \\
+\infty & |{\bf r}^+_i-{\bf r}^-_i| > R_c
\end{cases}
\label{vp}
\end{equation}
The pairwise potential energy $v^{pp}(\{{\bf r}^+,{\bf r}^- \}_i,\{{\bf r}^+,
{\bf r}^- \}_j)$ corresponds to an steric hindrance condition between two
associated pairs \cite{Stillinger}, since the distances between two unlike
ions corresponding to different pairs cannot be shorter than the minimum
distance between the ions that compose each pair:
\begin{equation}
v^{pp}(\{{\bf r}^+,{\bf r}^- \}_i,\{{\bf r}^+,{\bf r}^- \}_j)=
\begin{cases} +\infty & |{\bf r}^\mp_i-{\bf r}^\pm_j| \le d_{ij} \\ \\
0 & \text{otherwise}
\end{cases}
\label{vpp}
\end{equation}
with $d_{ij}\equiv min(|{\bf r}^+_i-{\bf r}^-_i|,|{\bf r}^+_j-{\bf r}^-_j|)$.

The interaction between the associated pairs and the free ions is modified by
the term $v^{p\pm}(\{{\bf r}^+,{\bf r}^- \}_i,{\bf r}^{\pm}_j)$,
that prevents the free ion to be closer to the unlike
ion of the associated pair than its partner:
\begin{equation}
v^{p\pm}(\{{\bf r}^+,{\bf r}^- \}_i,{\bf r}^{\pm}_j) =\begin{cases}
+\infty & |{\bf r}^\mp_i-{\bf r}^\pm_j| \le |{\bf r}^+_i-{\bf r}^-_i| \\ \\
 0 & \text{otherwise}
\end{cases}
\label{vp+-}
\end{equation}

Finally, two unlike free ions cannot be at a shorter distance than $R_c$ due to
the term $v^{+-}({\bf r}^+_i,{\bf r}^-_j)$:
\begin{equation}
v^{+-}({\bf r}^+_i,{\bf r}^-_j) = \begin{cases}
+\infty & |{\bf r}^+_i-{\bf r}^-_j| \le R_c \\ \\
0 & |{\bf r}^+_i-{\bf r}^-_j| > R_c
\end{cases}
\label{v+-}
\end{equation}

It is not hard to see that a mixture of associated pairs and free
ions with a potential energy given by the Eq. (\ref{hameffec}) is
completely equivalent to the original ionic system.  It is
interesting to note that, except the term on $v_p$ that ties the
ions that compose an associated pair, the other terms are
pairwise, short-ranged modifications of the hard-core conditions,
so the potential energy of an allowed mixture configuration is the
same as in the corresponding ionic configuration. Moreover, every
configuration of associated pairs and free ions obtained from an
ionic configuration by the pairing procedure described above
fulfil the constrains induced by the added potentials. Conversely,
the mixture configuration is the same as the one obtained from the
corresponding ionic configuration by the Stillinger-Lovett
protocol.

This analysis provides an exact chemical representation of the
ionic system as a mixture of bound pairs and free ions in chemical
equilibrium. The equivalence between representations of the system
is not only at the level of thermodynamic properties, but also in
the microscopic structure, unlike previous studies based on the
matching of the ``chemical'' and the ``physical'' free energies,
e.g. via the virial coefficients \cite{Hill,Zuckerman}. In those
studies the effective potentials given by Eqs. (\ref{vp}) and
(\ref{v+-}) could be guessed, but the other terms (that arise from
matching third and higher order virial coefficients) were not
clearly identified. Furthermore, they have not being used at all
in pairing theories of electrolytes. We must stress the importance
of the effective new term given by Eq. (\ref{vpp}) in the
potential energy, that reduces the configuration space available
to the associated pairs. This effect is specially important for
high values of $R_c$. An extreme case case that illustrates the
effect of missing the steric hindrance term is the loosely
tethered dimer fluid, in which the ions that compose the pair can
be at any relative distance greater than $\sigma_\pm$. In this
case, the canonical partition function of the dimer fluid is
$N_p!\times Z_U(N_p,N_p)$, where $N_p$ is the number of dimers,
$Z_U(N_p,N_p)$ is given by Eq. (\ref{partition2}) and the $N_p!$
factor is the number of different ways of pairing the unlike ions
in each ionic configuration and lead to different dimer
configurations. As a consequence, in the thermodynamic limit the
Helmholtz free energy per particle $f_d$ of the loosely tethered
dimer fluid diverges as:
\begin{equation}
\frac{f_d}{k_B T} \sim -\ln N_p + \left(\frac{2 f_i}{k_B T} + 1\right)
\end{equation}
where $f_i$ is the ionic Helmholtz free energy per ion, that it is well-behaved
in the case of neutral systems \cite{Lieb}.

The association degree of the ionic fluid in the coexistence region has been
studied by computer simulation in the framework of the previous analysis.
In the usual grand-canonical simulations, we have identified the associated
pairs by the Stillinger-Lovett rule with $R_c=\sqrt{3}L/2$, i.e. all the ions
are associated, and the minimum image convention is used in order to calculate
the distances between unlike ions. The probability distributions $p(r)$ of
having a pair an internal separation distance $r$ between ions is plotted
in the Fig. \ref{fig8} for the gas and liquid phases at $T^*=0.0443$.
Surprisingly, both distributions are practically identical, despite the
fact that the corresponding densities are quite different. The distributions
show a very pronounced peak for $r=\sigma_\pm$, and a local maximum
around $r=2.5\sigma_\pm$, approximately where $g_{+-}$ presents the second
local maximum. For large values of $r$, the distributions $p(r)$ decay to zero.
These results confirm the strong association in the ionic fluid at low
temperatures, and that the structure is weakly affected by variations in
density, at least in the range in which the gas-liquid coexistence occurs.
For values of $r$ close to the contact value, the distribution is well
described by the non-interacting pair fluid probability distribution,
given by the following expression:
\begin{eqnarray}
p^{ideal}(r)&\approx& \frac{r^2 \exp\left(\frac{q^2}{Dk_B T r}\right)}
{\int_{\sigma_\pm}^{R_c} y^2 \exp\left(\frac{q^2}{Dk_B T y}
\right)dy} \nonumber\\
\sim \frac{q^2 r^2}{D k_B T \sigma_\pm^4} &\times&\exp\left(\frac{q^2}
{Dk_B T\sigma_\pm}\left[\frac{\sigma_\pm}{r}-1\right]\right)
\end{eqnarray}
valid for $T^*\ll 1$ and an internal cutoff $R_c \gg \sigma_\pm$.
However, for $r \gtrsim 1.5\sigma_\pm$, $p(r)$ deviates from the
ideal expression as a consequence of the interaction between
associated pairs. The local maximum showed by the distribution
function indicates that the most bound pairs are solvated by less
bound pairs, to form stable neutral tetramers (and higher order
clusters, but their inclusion does not seem to affect the
conclusions of this Section). This fact is in agreement with the
features observed in the pair correlation functions in the
previous Section. We must stress that the structure that $p(r)$
presents is only due to the Coulombic interactions.

It is instructive to compare our results in the coexistence region
to the higher temperature ones. We have computed $p(r)$ at $T^*=1$
and the same range of densities (see Fig. \ref{fig9}). The
qualitative behavior of $p(r)$ at high temperatures is similar to
that predicted by Stillinger and Lovett \cite{Stillinger2}. First,
it shows a maximum localized in $r \sim \rho^{-\frac{1}{3}}$. On
the other hand, $p(r)$ takes significant values in a wider range
than the corresponding functions at lower temperatures, decaying
for large values of $r$ as $r^{-n}$, with $n\approx 3$. The latter
prediction differs from the value $n=4$ given in Ref.
\cite{Stillinger2} and requires further study to elucidate it.

The probability distribution $p^{R_c}(r)$ of having a pair an internal
separation distance between ions when the cutoff distance to define a
pair is equal to $R_c$ can be obtained from $p(r)$ as:
\begin{equation}
p^{R_c}(r)=\frac{p(r)}{\wp(R_c)}=\frac{p(r)}{\int_{\sigma_\pm}^{R_c} p(t)dt}
\end{equation}
where $\wp(r)\equiv \int_{\sigma_\pm}^r p(t)dt$ is the cumulant
distribution corresponding to $p(r)$. The density of associated
pairs is equal to $\rho \wp(R_c)/2$, and the total density of free
ions is $\rho [1-\wp(R_c)]$, where $\rho$ is the total density of
ions in the physical picture. It can be seen from Fig. \ref{fig8}
that for $R_c\gtrsim 3\sigma_\pm$, more than 95\% of ions are
associated into pairs in both vapor and liquid phases. So, if the
associated pair fluid phase separate in the same $(T,\rho)$ region
as the ionic fluid, it is expected than the free ions do play a
mere perturbative role in the phase coexistence. Actually, the
free ionic subsystem is a non-additive binary charged hard-sphere
mixture, where $R_c$ plays the role of the unlike-ion collision
diameter. However, the interactions between like ions are purely
repulsive, and it is not expected differences between the behavior
of this mixture and the RPM as soon as $\sigma_+$ and $\sigma_-$
are less than $R_c$. Furthermore, this system is in a polar
environment given by the associated pair subsystem, and
consequently the effective dielectric constant $D_{eff}$ of the
background will be increased. Then the free ion subsystem
is expected to have the same behavior as the RPM at the effective
reduced temperature $\tilde{T}$:
\begin{equation}
\tilde{T}=\frac{D_{eff} k_B T R_c}{q^2}=\frac{D_{eff}}{D} \times
\frac{R_c}{\sigma_\pm}\times T^*
\end{equation}
where $T^*$ is the reduced temperature. For $R_c \gtrsim 3\sigma_{\pm}$, the
critical temperature reduces at least to one third of the RPM reduced
critical temperature. Consequently, any possible free ion-driven vapor-liquid
transition should happen at much lower temperatures, and thus the free ion
subsystem should play no role in the vapor-liquid transition. In order to
check this hypothesis, the phase diagram of the associated
pair fluid has to be obtained. This issue will be addressed in the next
Section.

\section{Phase Behavior of the Associated Pairs Fluid}
\label{sec4}

The results obtained in the previous Section suggest that the ``chemical''
picture introduced above is a very convenient description of the ionic
fluid structure. Furthermore, we can analyse the role played by the associated
pairs in the phase equilibrium by eliminating the free ions. We have
performed grand-canonical Monte Carlo simulations of the associated dimer
system for $\lambda=3$ and different values of $R_c$. No free ions are allowed
($N_f\equiv 0$) and the potential energy of one configuration is given by
the Eq. (\ref{hameffec}). The grand-canonical free energies corresponding 
to these systems will consequently provide an upper bound to the ionic fluid 
grand-canonical free energy at the same temperature and pair chemical potential.
As for the ionic system, a fine-discretization approach with a refinement
parameter $\zeta=10$, and Ewald summation technique with conducting boundary
conditions is used to take into account the long range character of the
Coulombic interations. The basic steps are either insertion or deletion
of associated pairs (chosen randomly with the same probability),
biased with a Boltzmann distribution that depends on
the separation between the ions that compose the dimer. An associated pair
is inserted in the following way: the negative ion is placed at a random
place in the box, and its counterion is placed at a relative position
$\mathbf{r}_\pm$ ($\sigma_\pm \le |\mathbf{r}_\pm| < R_c$) following a
probability distribution proportional to $\exp\left(\beta_0 q^2
\phi_0(\mathbf{r}_\pm)\right)$, where $-\phi_0(\mathbf{r})$ is the Ewald
potential energy between two unlike ions of unit charge. On the other hand,
a pair is deleted with a probability proportional to $\exp\left(\beta_0
q^2 \phi_0(\mathbf{r}_\pm)\right)$, being $\mathbf{r}_\pm$ the relative
position of the positive ion of the pair with respect to the negative one.
The value of $\beta_0$ has been adjusted to improve the sampling during
the simulation. In this work we have set $\beta_0=8$.

The acceptance probabilities of a pair insertion $W_{ij}^i$ and a
pair deletion, $W_{ij}^d$ are the following:
\begin{widetext}
\begin{eqnarray}
W_{ij}^{i}&=&\textrm{min}\left(1,
(L^*)^3 \exp(\beta \mu - \beta \Delta U)
\left[\sum_{\sigma_\pm\le |\mathbf{r}_\pm|<R_c}
\frac{\exp(\beta_0 q^2\phi_0(\mathbf{r}_\pm))}{\zeta^3}\right]
\left[\sum_{k=1}^{N_p + 1} \exp(\beta_0
q^2\phi_0(\mathbf{r}_\pm^k))\right]^{-1} \right)\\
W_{ij}^{d}&=&\textrm{min}\left(1,
\frac{\exp(-\beta \mu - \beta \Delta U)}{(L^*)^3}
\left[\sum_{\sigma_\pm\le |\mathbf{r}_\pm|< R_c}\frac{\exp(\beta_0 q^2\phi_0
(\mathbf{r}_\pm))}{\zeta^3} \right]^{-1}
\left[\sum_{k=1}^{N_p} \exp(\beta_0
q^2\phi_0(\mathbf{r}_\pm^k))\right]\right)
\end{eqnarray}
\end{widetext}
where $i$ and $j$ are the initial and final configurations, respectively,
$N_p$ is the number of pairs at the configuration $i$, $\beta=1/(k_B T)$,
$\mu$ is the configurational chemical potential, and $\Delta U$ is the
``chemical'' potential energy variation in the movement, including the
steric hindrance terms given by Eq. (\ref{vpp}). As it can be seen,
the acceptance probabilities reduce to the usual ones as $\beta_0 \to 0$.

Effective critical points for different values of $L^*$ were
estimated by using mixed-field finite-size scaling methods
\cite{Bruce} and assuming Ising-like criticality. Although recent
results \cite{Fisher2} indicate that the pressure should also
enter the field mixing, our approach should be satisfactory to
discern the dependence of the critical parameters on $R_c$.
Moreover, a systematic study for the RPM case has shown that the
assumed universality class is very likely to be the right one
\cite{Luijten}. We use histogram reweighting techniques
\cite{Ferrenberg} to combine the histograms from different runs
(typically three) and estimate the critical parameters and their
standard errors. Very long runs are needed in order to overcome
the critical slowing down and the low acceptance ratios due to the
low temperatures involved. Defining a step as a try of a pair
insertion/deletion, we have performed simulations of about
$2\times 10^7-4\times 10^7$ equilibration steps and $2\times
10^8-4\times 10^8$ sampling steps for $L^*=12$; $5 \times
10^7-2\times 10^8$ equilibration steps ans $6\times 10^8 -
1.2\times 10^9$ sampling steps for $L^*=15$; and $2\times 10^8$
equilibration steps and $1.2\times 10^9 - 1.8\times 10^9$ sampling
steps for $L^*=18$, in order to get good statistics for the
histograms. The (effective) critical parameters are obtained by
minimizing the deviation of the appropriately scaled mixed-field
${\cal M}\propto \rho-s u$ marginal probability distribution ($u$
is the potential energy density and $s$ is the mixing parameter)
with respect to the corresponding critical 3-dimensional Ising
universal function \cite{Bruce,Thanos3}. As also occurs in the
ionic and tethered dimer systems, the matching improves as $L^*$
increases (see Fig. \ref{fig10}). Our estimations of the critical
parameters for different values of $L^*$ and $R_c^*\equiv
R_c/\sigma_\pm$ are listed in Table \ref{tablecritpar} and plotted
in Figs. \ref{fig11} and \ref{fig12}. The smallest value of
$R_c^*$ is the same as the maximum allowed separation in the
tethered dimer system, and as expected the results obtained for
the associated pairs fluid are practically indistinguishable from
the tethered dimer system. As $R_c^*$ increases, it is observed a
sharp increase in the critical temperature but the critical
density remains almost unaffected. However, for $R_c^*\approx 3-4$
the critical temperature reaches a maximum and then decreases,
converging towards the free ion critical temperature. On the other
hand, the critical density also decreases towards the free ion
critical density, although the statistical uncertainties are
bigger for this parameter and the exact path of convergence is
less defined. We must also point out that the statistical
uncertainties are big enough not to make possible an estimation of
correction-to-scaling effects.

For an heuristic explanation for such a behavior, we have to consider
the energetically favoured configurations. It is important to consider
the interaction between two different associated pairs,
that has its minimum energy configuration in a square ring
conformation for low values of $\lambda$, with a close-energy secondary
minimum conformation for the linear $(+,-)$ chain \cite{E2PM}.
If $R_c$ is too small, by deforming the square it is possible to find
energetically relevant configurations (i.e. the potential energy difference
per ion of such configurations with respect to the minimum one is of
the same order as the thermal energy) in which one of the associated
pairs has an internal ionic separation less than $R_c$, but the other
one does not fulfil such a condition. Consequently, a greater value of
$R_c$ also increases the number of energetically relevant configurations
between two associated pairs. Then the effective interaction between
associated pairs is enhanced, and thus an increase of the critical
temperature should be expected until all the relevant configurations
are allowed (that occurs for $R_c \sim 3\sigma_\pm$). On the other
hand, the average size of the associated pairs and their aggregates is
increased, leading to a decrease on the critical density (in order to keep
the reduced density in terms of the effective particle size constant).
These arguments are not longer valid for $R_c^*\gtrsim 3$. Our results show
the inclusion of associated pairs with larger internal ionic separations
are not crucial to the gas-liquid phase transition but also that they
bar the phase transition, as it can inferred from the critical
parameters decrease.

As for the ionic and tethered dimer models, histogram reweighting
techniques \cite{Ferrenberg,Thanos3} allow us to obtain
the coexistence curve up to $T \lesssim 0.98 T_c$. As the finite-size
effects are not important far from the critical point, we have considered
$L^*=15$. Taking advantage of the wide range of densities covered by
near-critical histograms, we combine them with liquid subcritical state
histograms in order to extend the density range. The extra simulations
involve shorter runs (typically $2\times 10^8$ steps after equilibration).
The gas-liquid coexistence curves for different values of $R_c^*$
are plotted in Fig. \ref{fig13}, showing the same tendency as the observed
one in the critical parameters. It is interesting to note that the vapor
branches coincide (except close to the critical point) for $R_c > 3\sigma_\pm$.
We conclude from this observation that is in the liquid branch where the
ionic fluid differs mostly from the associated pair system.

\section{Discussion and Conclusions}

In summary, an exact ``chemical'' representation of the ionic
system as a mixture of $(+,-)$ associated ion pairs and free ions
has been introduced. This representation, closely related to the
Stillinger-Lovett pairing procedure for the RPM, has the advantage
that not only provides an exact matching of between the physical
and chemical representation thermodynamics, but also from a
microscopic point of view.  It also avoids an entropy catastrophe
that occurs for the tethered dimer model studied at large values
of the tether length.  The addition to the physical hamiltonian of
some new pairwise hard-core interactions between the ``chemical''
components provides the suitable hamiltonian for the ``chemical''
representation. In the low temperature and low density regime in
which the ionic vapor-liquid transition occurs, such a
representation provides a faithful characterization of the
microscopic structure of the ionic fluid, and it can be the basis
of new theoretical approaches.

The analysis of the phase behavior of the system only composed by
associated pairs indicates strongly that the ionic fluid
vapor-liquid transition is driven by them. For $R_c\lesssim 3
\sigma_\pm$, the critical temperature increases with $R_c$, as
more energetically favoured configurations are allowed. For
$R_c\gtrsim 3\sigma_\pm$, the critical parameters converge
smoothly \emph{from above} towards the ionic fluid critical
parameters as $R_c$ increases. This fact indicates not only that
the free ions do not drive the phase transition, but also have the
opposite effect in the transition. The value of $R_c\approx
3\sigma_\pm$ corresponding to the maximum on the critical
temperature, can be regarded as the optimal size of the associated
pair.

Some remarks on the limitations of the present work are
appropriate at this point. We have focussed only on the effect of
the associated pairs in the thermodynamical properties. However,
the conducting character of the ionic fluid is completely driven
by the free ions. On the other hand, the remarkable success of
pairing theories (even when the ``chemical'' representation they
implicitly use is not completely correct) remains unexplained. It
is possible that the associated pair solvation, in addition to the
ion-ion correlations, can mimic the pair-pair interactions.
Further studies are needed to completely solve the origin of the
ionic fluid vapor-liquid phase transition.

\begin{acknowledgments}
J.M.R.-E. and L.F.R. gratefully acknowledge financial support for
this research by Grant No. PB97-0712 from DGICyT (Spain) and No.
FQM-205 from PAI (Junta de Andaluc\'{\i}a). J.M.R.-E. also wishes
to thank an FPI scholarship from Ministerio de Educaci\'on y
Cultura (Spain). AZP acknowledges financial support from the
Department of Energy (grant DE-FG02-01ER1512) and ACS-PRF (grant
38165-AC9).
\end{acknowledgments}

\clearpage
\begin{table} [t]
\caption {\label{tablecritpar} Dependence on $R_c^*$ of the estimated
critical parameters. (The 1 $\sigma$ statistical uncertainties refer to
the last decimal places.)}
\vspace{1mm}
\begin{ruledtabular}
\begin{tabular}{ccclc}
$R_c^*$ & $L^*$ & $T_c^{*}\times 10^2$ & ~~~$-\mu_c^*$
& $\rho_c^{*}\times 10^2$\\
\hline
~1.02 & 12 & 4.24(2) & 1.2512(4)  & 7.66(12) \\
     & 15 & 4.25(1) & 1.2504(2) & 7.56(9)~~\\
     & 18 & 4.26(1) & 1.25077(8) & 7.34(10) \\
1.5& 12 & 4.44(1) & 1.30489(10) & 7.41(12) \\
     & 15 & 4.46(1) & 1.30565(5) & 7.52(10) \\
     & 18 & 4.47(1) & 1.30580(8) & 7.33(17) \\
2.0& 12 & 4.53(2) & 1.3112(3)& 7.45(14) \\
     & 15 & 4.53(1) & 1.3109(2) & 7.53(18) \\
     & 18 & 4.53(2) & 1.3107(3)& 7.54(17) \\
3.0& 12 & 4.65(2) & 1.3187(4)& 7.23(10) \\
     & 15 & 4.60(1) & 1.31792(13)& 6.93(12) \\
     & 18 & 4.62(3) & 1.3182(4)& 6.69(10) \\
4.0& 12 & 4.56(4) & 1.3202(5)& 7.05(22) \\
     & 15 & 4.59(2) & 1.3207(5)& 6.45(14) \\
     & 18 & 4.62(3) & 1.3213(5)& 6.4(2)~~~ \\
5.0& 12 & 4.58(1) & 1.32210(14)& 6.17(21) \\
     & 15 & 4.57(1) & 1.32230(7)& 6.46(24) \\
     & 18 & 4.56(4) & 1.3220(7)& 6.17(10) \\
6.0& 12 & 4.55(1) & 1.3222(3)& 6.41(2)~~ \\
     & 15 & 4.51(2) & 1.3214(3)& 6.33(10) \\
     & 18 & 4.59(3) & 1.3222(6)& 6.1(2)~~~ \\
7.0& 15 & 4.51(2) & 1.3216(5)& 6.33(10) \\
     & 18 & 4.50(2) & 1.3211(3)& 6.04(17) \\
8.0& 18 & 4.53(4) & 1.3224(6)& 6.1(4)~~~ \\
9.0& 18 & 4.49(1) & 1.3215(3)& 5.6(3)~~~ \\
Free ion case \footnote{It is equivalent to consider $R_c^*\equiv
\sqrt{3}L^*/2$.}& 12 & 4.49(1) & 1.3200(1)& 6.37(7)~~\\
& 15 & 4.48(1) & 1.3199(1)& 6.04(11) \\
& 18 & 4.49(2) & 1.3199(4)& 6.05(11) \\
\end{tabular}
\end{ruledtabular}
\end{table}
\clearpage
\clearpage
\begin{figure}
\includegraphics[width=8.5cm]{fig3.eps}
\caption{The same as the Fig. \ref{fig2} in the liquid branch: $T^*=0.0425$
and $\rho^*=0.162(3)$.\label{fig3}}
\end{figure}
\begin{figure}
\begin{center}
\includegraphics[width=8.5cm]{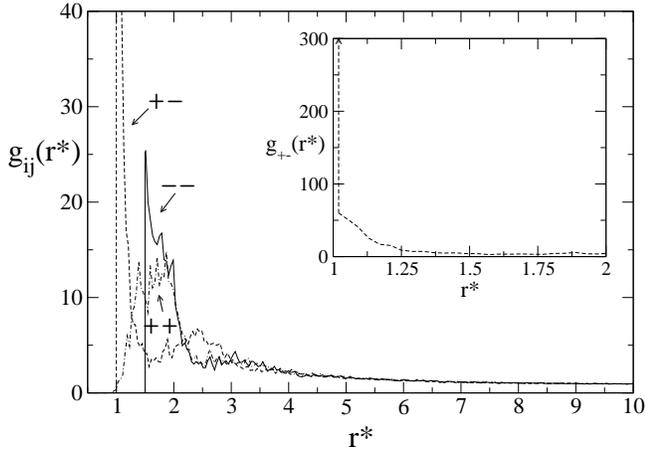}
\end{center}
\caption{Ion-ion radial distribution functions corresponding to the tethered
dimer system at $T^*=0.0405$ and $\rho^*=0.0119(15)$ (the gas phase).
In the inset, the behavior of $g_{+-}(r^*)$ at distances close to
the contact value.\label{fig4}}
\end{figure}
\clearpage
\begin{figure}
\includegraphics[width=8.5cm]{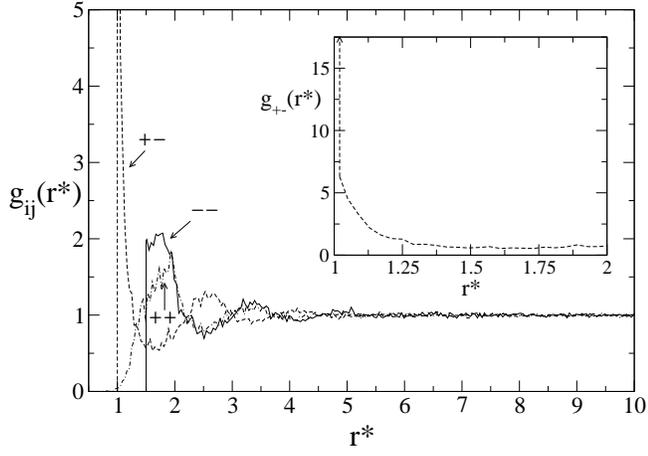}
\caption{The same as the Fig. \ref{fig4} in the liquid branch: $T^*=0.0405$
and $\rho^*=0.206(2)$.\label{fig5}}
\end{figure}
\begin{figure}
\includegraphics[width=8.5cm]{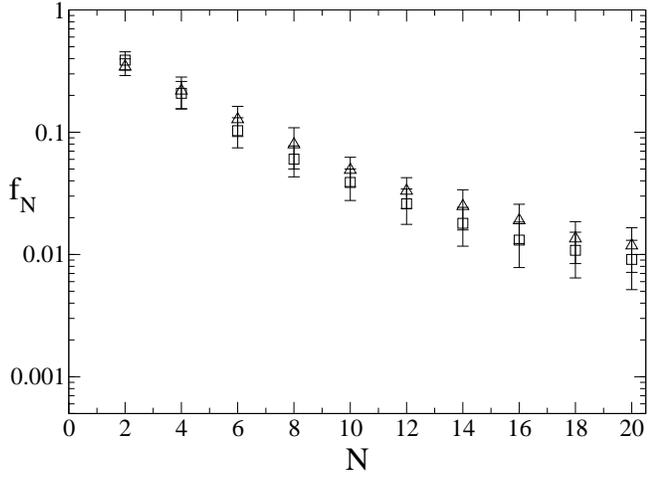}
\caption{Fraction $f_N$ of $N$-ion neutral clusters. The squares
correspond to the ionic fluid at $T^*=0.0425$ and $\rho^*=0.0076(15)$,
and the triangles to the tethered dimer fluid at $T^*=0.0405$ and
$\rho^*=0.0119(15)$.\label{fig6}}
\end{figure}
\clearpage
\begin{figure}
\includegraphics[width=8.5cm]{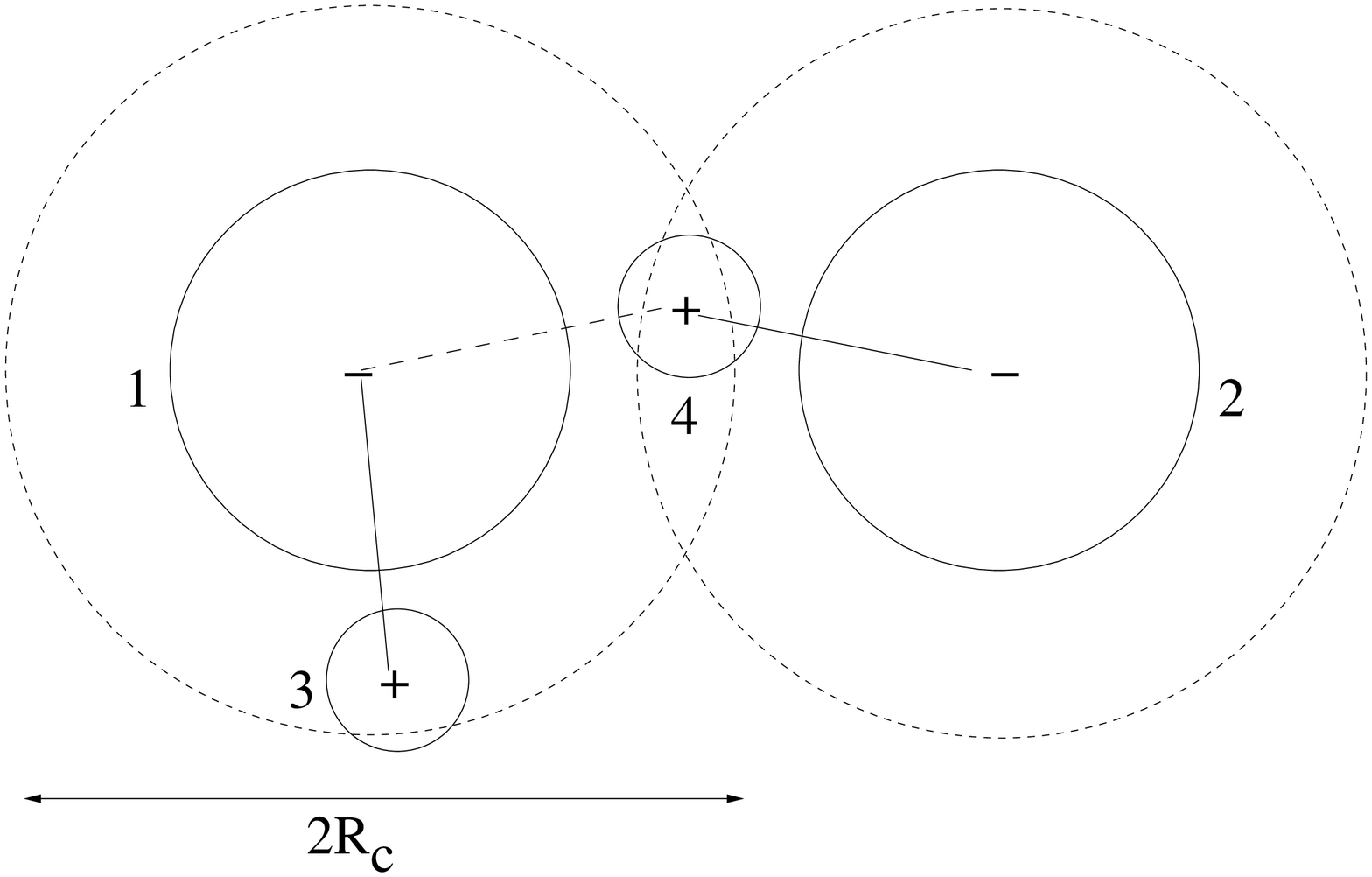}
\caption{Different ways of pairing ions using the criterion that two unlike
ions are paired if the distance between them is shorter than the cutoff $R_c$.
The cation 4 can be paired with either the anion 1 or the anion 2. In the
first case the ions 2 and 3 remain free, and in the latter case the
ions 1 and 3 constitute another associated pair.\label{fig7}}
\end{figure}
\begin{figure}
\includegraphics[width=8.5cm]{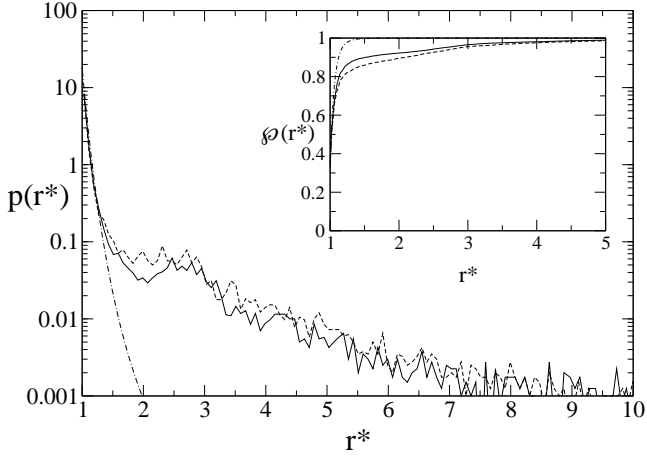}
\caption{Probability distribution $p(r^*)$ of having a pair an internal
separation distance $r^*$ between ions at $T^*=0.0443$ and reduced densities
$\rho^*=0.0064(7)$ (solid line), corresponding to the gas phase, and
$\rho^*=0.149(3)$ (dashed line), corresponding to the liquid phase.
The dot-dashed line corresponds to the ideal dimer fluid approach
(see text). In the inset, the cumulant distribution $\wp(r^*)\equiv
\int_1^{r^*}p(t)dt$. The meaning of the symbols is the same as before.
\label{fig8}}
\end{figure}
\clearpage
\begin{figure}
\includegraphics[width=8.5cm]{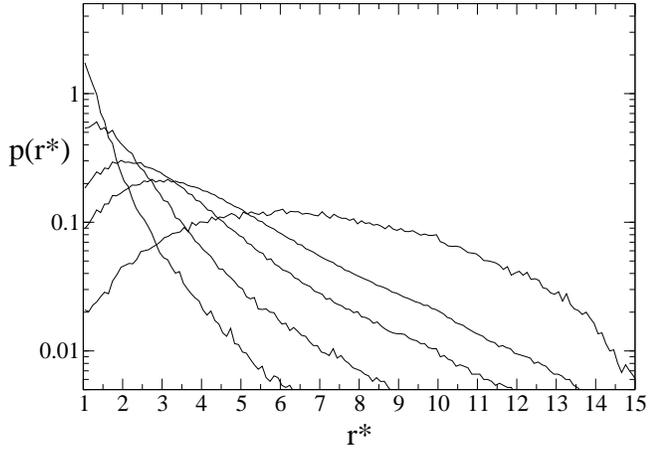}
\caption{Probability distribution $p(r^*)$ at $T^*=1$. At $r^*=1$, from bottom
to top the lines correspond to reduced densities $\rho^*=0.00113(1),0.00618(1),
0.01453(2), 0.0456(3)$ and $0.1566(6)$.
\label{fig9}}
\end{figure}
\begin{figure}
\includegraphics[width=8.5cm]{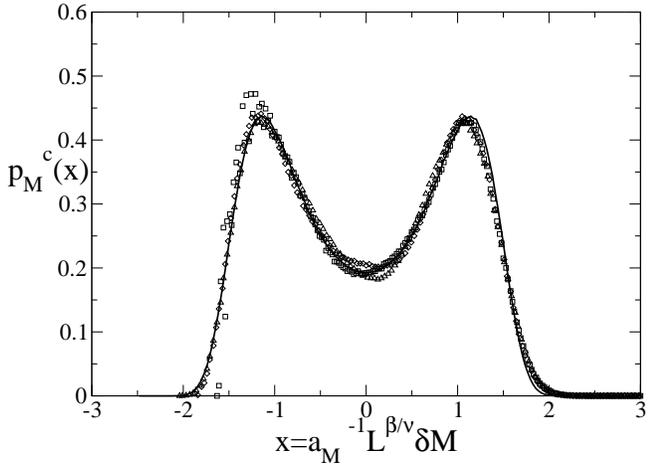}
\caption{Rescaled marginal probability distributions $p_{\cal M}^c(x)$ for
the associated pairs fluid characterized by $\lambda=3$ and $R_c^*=3$.
The solid line is the universal function corresponding to the 3-dimensional
Ising universality class, and the symbols correspond to the best-matching
simulation results: the squares correspond to $L^*=12$, the diamonds to
$L^*=15$ and the triangles to $L^*=18$.
\label{fig10}}
\end{figure}
\clearpage
\begin{figure}
\includegraphics[width=8.5cm]{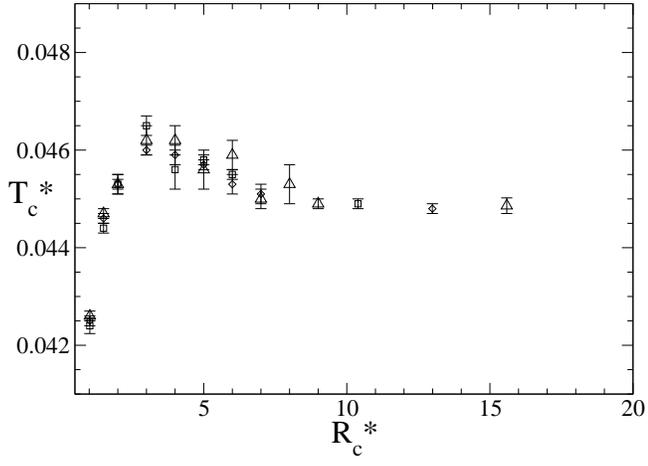}
\caption{Dependence of the critical temperature $T_c^*$ on the reduced
cutoff distance $R_c^*$, for $L^*=12$ (squares), $L^*=15$ (diamonds)
and $L^*=18$ (triangles).
\label{fig11}}
\end{figure}
\begin{figure}
\includegraphics[width=8.5cm]{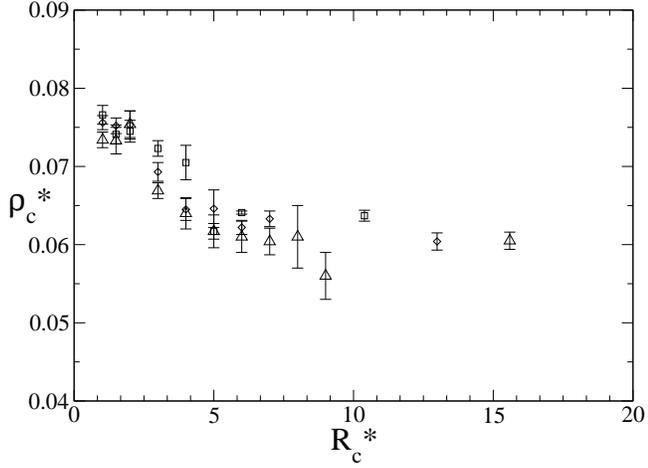}
\caption{Dependence of the critical density $\rho_c^*$ on the reduced
cutoff distance $R_c^*$, for $L^*=12$ (squares), $L^*=15$ (diamonds)
and $L^*=18$ (triangles).
\label{fig12}}
\end{figure}
\clearpage
\begin{figure}
\includegraphics[width=8.5cm]{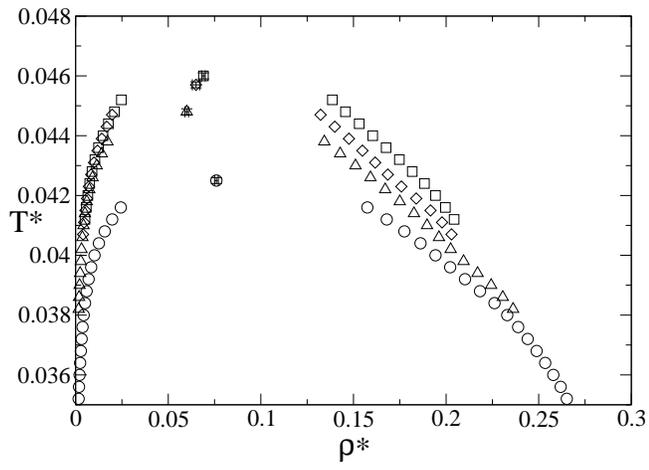}
\caption{Gas-liquid coexistence curves for $R_c^*=1.02$ (circles),
$R_c^*=3$ (squares), $R_c^*=5$ (diamonds) and the free ion system
(triangles).\label{fig13}}
\end{figure}
\end{document}